\documentclass[aps,prl,reprint,superscriptaddress,floatfix,longbibliography]{revtex4-2}

\usepackage[T1]{fontenc}
\usepackage[utf8]{inputenc}
\usepackage{amsmath,amssymb,amsfonts,bm,mathtools}
\usepackage{graphicx}
\usepackage{booktabs}
\usepackage{microtype}
\usepackage{xcolor}
\usepackage{hyperref}
\usepackage{enumitem}

\hypersetup{colorlinks=true,linkcolor=blue,citecolor=blue,urlcolor=blue}
\graphicspath{{./}}
\allowdisplaybreaks

\newcommand{\ket}[1]{|#1\rangle}

\newcommand{\CP}{\mathrm{CP}}
\newcommand{\CZ}{\mathrm{CZ}}
\newcommand{\CH}{\mathrm{CH}}
\newcommand{\CNOT}{\mathrm{CNOT}}
\newcommand{\HCNOTT}{\ensuremath{H\text{-}\CNOT\text{-}T}}

\newcommand{\GiB}{\mathrm{GiB}}
\newcommand{\qsim}{\texttt{qsim}}
\newcommand{\quimb}{\texttt{quimb}}
\newcommand{\cotengra}{\texttt{cotengra}}

\newcommand{\START}{\mathrm{START}}

\begin{document}

\title{Simulating quantum circuits with a neural statebank}

\author{Taige Wang}
\thanks{taige\_wang@fas.harvard.edu}
\affiliation{Materials Research Laboratory, Massachusetts Institute of Technology, Cambridge, MA 02139, USA}
\affiliation{Department of Physics, Harvard University, Cambridge, MA 02138, USA}

\author{Liang Fu}
\affiliation{Department of Physics, Massachusetts Institute of Technology, Cambridge, MA 02139, USA}

\date{\today}

\begin{abstract}
Predicting the output of quantum circuits is a central bottleneck for verifying quantum processors because a generic wavefunction grows exponentially with system size. We introduce a neural statebank that learns this wavefunction along the circuit trajectory. Each layer is stored as an autoregressive Transformer checkpoint trained from local gate updates to the preceding checkpoint, producing a compact neural representation that can evaluate amplitudes and generate independent samples. On long-range circuits combining entanglement, magic, and non-diagonal branching, a 0.3-million-parameter statebank reaches $\sim 10^{-2}$ infidelity at 34 qubits, outperforming the other tested approximate simulators while using far less memory than exact state-vector evolution. The same architecture accurately simulates quantum approximate optimization, Clifford+$T$, and Clifford circuits.
\end{abstract}

\maketitle


\section*{Introduction}

Quantum processors are reaching regimes in which their outputs are difficult to predict classically, yet such predictions remain essential for verification, debugging, algorithm design, error mitigation, and claims of quantum advantage~\cite{Feynman1982,Preskill2018,Arute2019,Temme2017,LiBenjamin2017,Cai2023}. The central obstacle is representation of the quantum state. For $N$ qubits, a generic wavefunction contains $2^N$ complex amplitudes, so exact state-vector evolution becomes memory-limited after only a few dozen qubits. Storing a complex64 state vector alone requires $8\times 2^N$ bytes, reaching $128\,\GiB$ already at $N=34$.

Classical simulation therefore depends on compression, but existing compression strategies work best in different favorable regimes. Tensor-network contraction avoids storing all amplitudes, but its cost grows with an effective contraction width~\cite{MarkovShi2008,GrayKourtis2021,Huang2021}. Matrix-product states and related ansatzes are efficient when entanglement is limited~\cite{Vidal2003,Vidal2004,Schollwock2011,VerstraeteCirac2004,Paeckel2019}. Stabilizer methods efficiently simulate Clifford circuits~\cite{AaronsonGottesman2004}, and near-Clifford methods extend this regime at a cost controlled by magic~\cite{BravyiGosset2016,Bravyi2019}. Observable-centered Heisenberg propagation methods can track selected expectation values when the observables are local, but they do not generally provide the wavefunction itself~\cite{Gottesman1998Heisenberg,RallLiangCookKretschmer2019PauliPropagation,AnandTemmeKandalaZaletel2023ZNEBenchmark,Angrisani2025NoiselessObservables}. These methods are powerful, but a generic circuit need not remain within any one favorable structure.

Neural wavefunctions offer another compression principle. Instead of storing an amplitude table, one learns a compact neural representation of the quantum state that assigns amplitudes to basis configurations~\cite{CarleoTroyer2017,GaoDuan2017,Sharir2020,HibatAllah2020,Vaswani2017}. A central obstacle is how to obtain the training data. A direct supervised strategy would require computing many final-state amplitudes in the computational basis before training the neural network. That is already the exponentially hard simulation problem~\cite{HouGarrattEassa2025ManyMeasurements,Aaronson2020ShadowTomography,HuangKuengPreskill2020ClassicalShadows,Elben2023RandomizedMeasurementToolbox}.

Here we introduce a neural statebank for quantum-circuit simulation. Instead of reconstructing the final wavefunction in one step, the method learns the circuit trajectory layer by layer. We decompose the circuit as $U=U_D\cdots U_2U_1$ and store a sequence of checkpoints $\theta_0,\theta_1,\ldots,\theta_D$, where $\psi_{\theta_\ell}$ approximates the state after the first $\ell$ layers. Starting from the known initial state, each checkpoint is trained from the preceding checkpoint using only the local action of the next circuit layer. The result is a sequence of compact neural representations that can evaluate amplitudes and generate samples from the learned Born distribution.

Earlier neural circuit simulations explored layerwise learning with neural wavefunctions~\cite{JonssonBauerCarleo2018NeuralNetworkStates}. The statebank turns this idea into a collection of directly sampleable neural representations, one for each intermediate circuit state. The Transformer architecture provides the expressive power needed to represent highly entangled states~\cite{Vaswani2017,ZhangDiVentra2023,Viteritti2023,Geier2025SelfAttention,Zaklama2025FoundationModel, Wang2026WalshExpressibility,Paul2026EntanglementBoundNQS}. The autoregressive factorization accelerates optimization by generating independent samples without a Metropolis chain~\cite{Sharir2020,HibatAllah2020,Luo2022ARTransformer,Barrett2022,IbarraGarciaPadilla2025AutoregressiveHubbard,Lange2024NQSReview}. This is especially useful for circuit states, because a circuit does not define a natural local proposal graph: a single spin flip often produces a basis state with very small probability, so a Metropolis chain can waste many amplitude evaluations on rejected proposals. Together, the autoregressive Transformer representation and the layerwise statebank protocol provide a neural simulator for circuits that combine entanglement, magic, and non-diagonal branching.

\begin{figure}[t]
    \centering
    \includegraphics[width=0.98\linewidth]{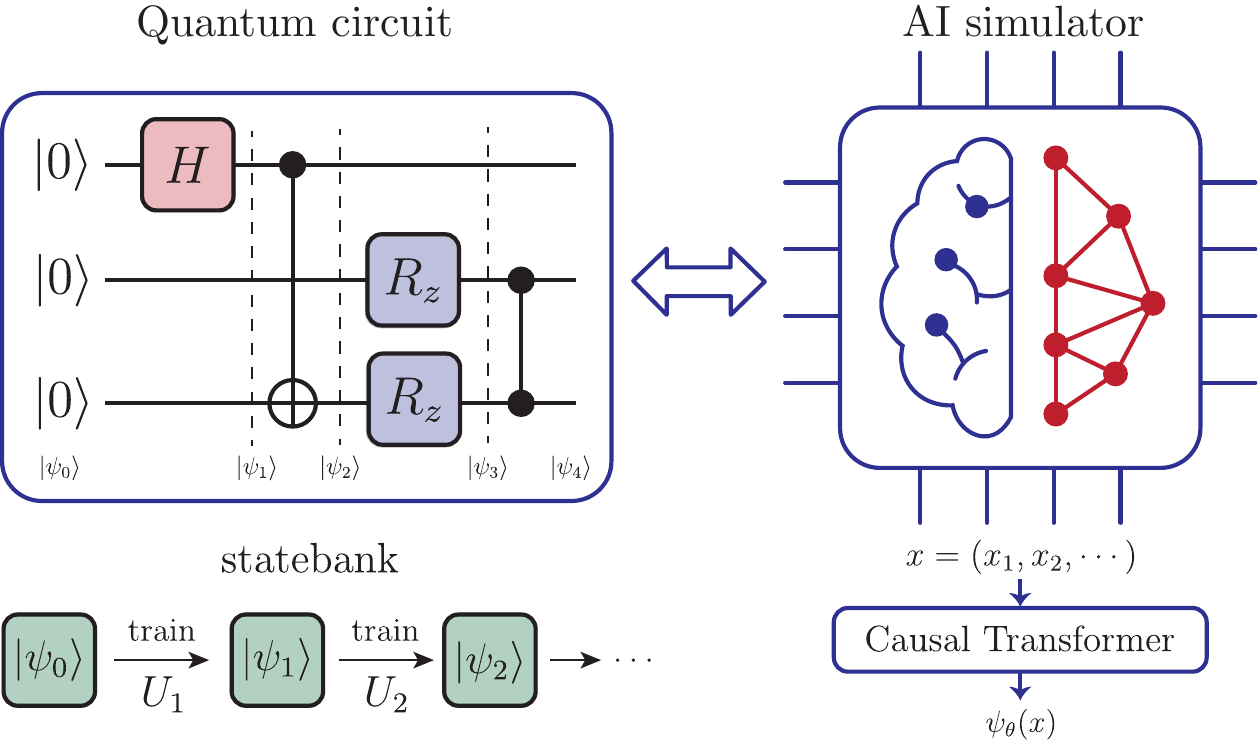}
    \caption{\textbf{Statebank training.}
A circuit is decomposed into layers $U_1,U_2,\ldots$, which produce intermediate wavefunctions $\ket{\psi_0},\ket{\psi_1},\ldots$. Instead of storing these states as full $2^N$-component vectors, the simulator represents each one by an autoregressive Transformer checkpoint. To train checkpoint $\theta_\ell$, samples and amplitudes from $\theta_{\ell-1}$ are combined with the exact local action of $U_\ell$ to define targets for the next state.}
    \label{fig:architecture}
\end{figure}

\section*{Neural statebank}

The neural statebank follows the circuit trajectory with a sequence of Transformer wavefunction checkpoints as shown in Fig.~\ref{fig:architecture}. The exact states obey
\begin{equation}
|\psi_\ell\rangle=U_\ell|\psi_{\ell-1}\rangle,
\qquad
|\psi_0\rangle \ \text{known}.
\end{equation}
The intermediate $2^N$-component wavefunctions are never stored. Instead, checkpoint $\theta_\ell$ defines a neural amplitude $\psi_{\theta_\ell}(x)$ that approximates $\psi_\ell(x)=\langle x|\psi_\ell\rangle$, where $x=(x_1,\ldots,x_N)$ is a computational-basis bitstring. At layer $\ell$, the previous checkpoint serves as the best available proxy for $\psi_{\ell-1}$. Applying the next circuit layer to this proxy defines an accessible teacher state, and a new checkpoint is trained to approximate that teacher. Thus $\psi_{\theta_\ell}$ learns $U_\ell\psi_{\theta_{\ell-1}}$, which tracks the exact state when the previous checkpoint is accurate.

Each checkpoint is an autoregressive Transformer neural quantum state. For a bitstring $x=(x_1,\ldots,x_N)$, we write
\begin{equation}
\psi_\theta(x)
=
\exp\left[
\frac{1}{2}\sum_{i=1}^{N}\log p_\theta(x_i|x_{<i})
+
i\sum_{i=1}^{N}\varphi_\theta(x_i|x_{<i})
\right].
\label{eq:nqs_ansatz}
\end{equation}
The Transformer reads the qubits sequentially with a causal attention mask, so the output at site $i$ depends only on the prefix $x_{<i}$. It predicts both the conditional probability for the next bit and the corresponding phase increment. This autoregressive structure supplies the neural Born distribution by construction and enables direct ancestral sampling of $x_1,x_2,\ldots,x_N$, without a Metropolis chain. It therefore avoids equilibration and autocorrelation costs that would otherwise slow down neural-wavefunction training for circuit simulation.

The remaining task is to generate targets without knowing the exact next state. The key simplification is locality. For a diagonal layer, the update is pointwise: each bitstring amplitude is multiplied by a known phase. For a non-diagonal gate $u$ acting on a small set $S$ of one or two qubits, all other qubits are spectators. We write the active bits as $\alpha=x_S$ and the spectator bits as $y=x_{\bar S}$. Since the gate acts trivially outside $S$, it cannot change $y$, and the target amplitude is obtained by varying only the active bits,
\begin{equation}
\widetilde\psi_\ell\bigl(y,\alpha\bigr)
=
\sum_{\beta\in\{0,1\}^{|S|}}
u_{\alpha\beta}\,
\psi_{\theta_{\ell-1}}\bigl(y,\beta\bigr).
\label{eq:local_gate_update}
\end{equation}
Here $\widetilde\psi_\ell=U_\ell\psi_{\theta_{\ell-1}}$ is the accessible teacher state, and $(y,\beta)$ denotes the full bitstring with the same spectator configuration and active bits $\beta$. In training, we sample spectator configurations from the previous checkpoint, query the locally related bitstrings, apply the exact gate matrix, and train the new checkpoint to match the resulting amplitudes. The statebank therefore learns the circuit trajectory from local gate rules and direct neural samples, rather than from an explicitly constructed wavefunction.

\begin{figure*}[t]
    \centering
    \includegraphics[width=\linewidth]{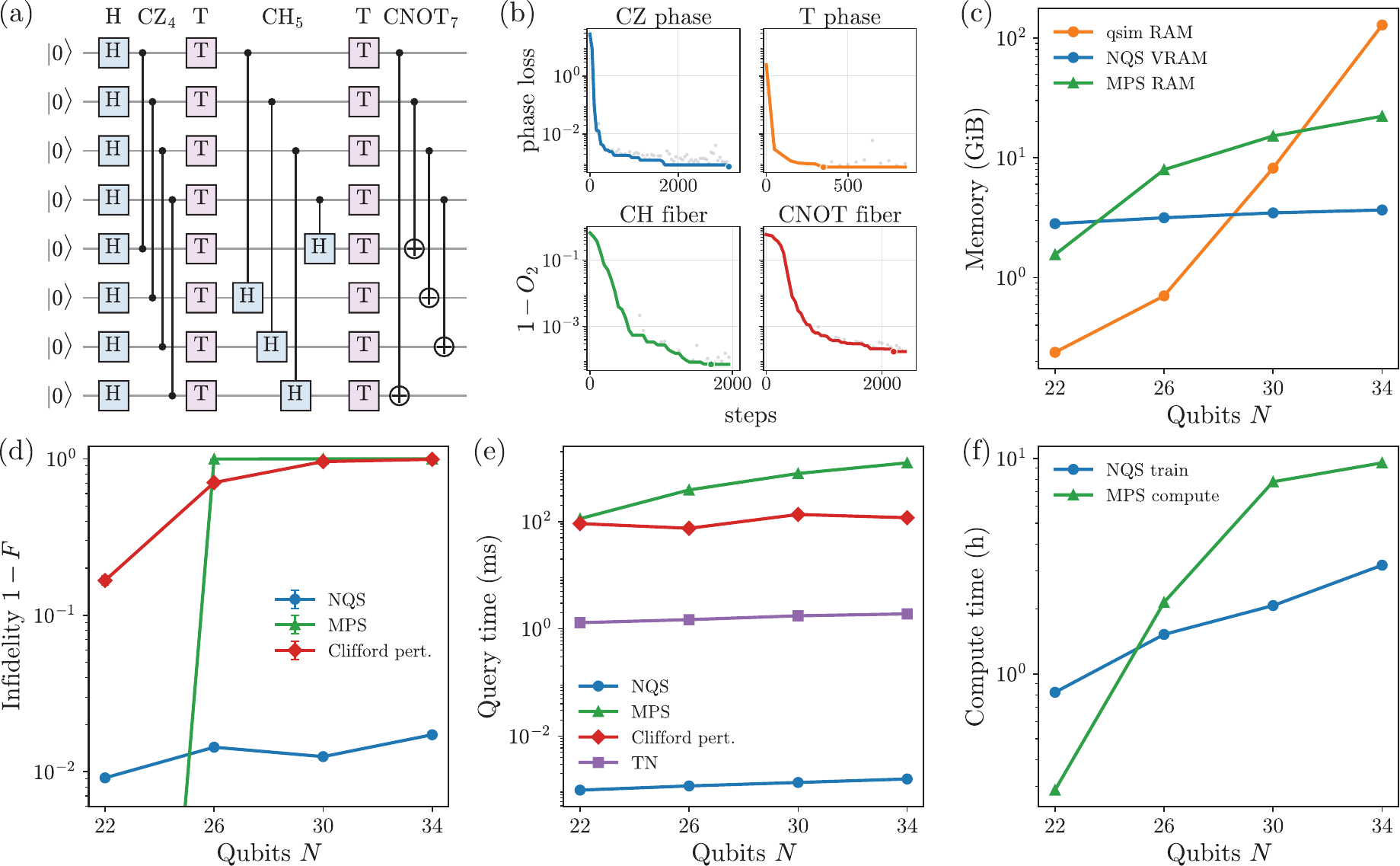}
    \caption{\textbf{HCZCH benchmark.}
\textbf{a}, HCZCH circuit shown for $N=8$ qubits. \textbf{b}, Representative layer-by-layer training diagnostics. Diagonal layers are monitored by phase error, whereas CH and CNOT layers are monitored by local-unitary overlap error. \textbf{c}, Memory use. Exact state-vector storage reaches $128\,\GiB$ at $N=34$, while the Transformer statebank remains near $3\,\GiB$ over the widths studied. \textbf{d}, Final infidelity. The statebank remains near $10^{-2}$ from $N=22$ to $34$, whereas MPS and near-Clifford perturbation deteriorate at larger width. \textbf{e}, Amplitude-query time. Tensor-network contraction is competitive for selected amplitudes, but the statebank gives fastest amortized queries after training. \textbf{f}, End-to-end compute time for producing learned state representations.}
    \label{fig:showcase}
\end{figure*}

\section*{HCZCH benchmark}

To test the statebank in a regime where several standard methods are stressed at once, we introduce a shallow but deliberately demanding circuit. The benchmark is the HCZCH circuit in Fig.~\ref{fig:showcase}(a),
\begin{equation}
H_{\rm all}
\rightarrow \CZ_4
\rightarrow T_{\rm all}
\rightarrow \CH_5\rightarrow T_{\rm all}
\rightarrow \CNOT_7.
\end{equation}
Here $G_s$ denotes a layer of two-qubit gates $G$ between qubits separated by offset $s$. Although the circuit has only five layers after the initial Hadamards, it combines several sources of classical difficulty: long-range $\CZ$ and CNOT layers create nonlocal entanglement, all-qubit $T$ layers introduce magic, and the CH layer mixes amplitudes between computational-basis strings.

The panels in Fig.~\ref{fig:showcase}(c--f) compare four standard simulation strategies on this circuit. State-vector evolution with \qsim{} is exact, but it requires the full $2^N$ memory, reaching the $128\,\GiB$ scale at $N=34$~\cite{GoogleQsim,Isakov2021qsim,Arute2019}. MPS simulation with \quimb{} is efficient when entanglement remains low, but the long-range entangling layers rapidly increase the required bond dimension. At bond dimension $\chi=3200$, MPS is essentially exact at $N=22$, whereas its infidelity becomes order one at larger sizes and its memory footprint reaches the tens-of-GiB scale~\cite{Vidal2003,Vidal2004,Schollwock2011,Paeckel2019,Gray2018quimb}. Near-Clifford perturbation exploits stabilizer structure but is stressed by magic: in this circuit, the two $T_{\rm all}$ layers together with the CH layer produce large infidelity even at perturbation rank four million, while the amplitude-query time reaches the $10^2\,{\rm ms}$ scale~\cite{AaronsonGottesman2004,BravyiGosset2016,Bravyi2019}. Tensor-network contraction with \cotengra{} remains practical for selected amplitudes, but it is a path-width-limited query method rather than a stored representation of the wavefunction~\cite{MarkovShi2008,Gray2018quimb,GrayKourtis2021,Huang2021}.

The neural statebank performs well in this mixed hardness regime. We deliberately use a small Transformer with $d_{\rm model}=128$, two Transformer blocks, four attention heads, feed-forward dimension $d_{\rm ff}=256$, and zero dropout. The model has only $270{,}468$ parameters at $N=34$. Fig.~\ref{fig:showcase}(b) shows the layer-by-layer training procedure: diagonal layers are learned by phase updates, while CH and CNOT layers are learned through active-qubit constraints. The resulting statebank uses only $\sim 3\,\GiB$ of memory, keeps the final infidelity near the $10^{-2}$ level from $N=22$ to $34$, and provides millisecond-scale amortized amplitude queries after training. This benchmark illustrates the central advantage of the method: it produces a compact neural representation of the wavefunction in a circuit where entanglement, magic, and non-diagonal branching stress different specialized simulators at the same time.

\begin{figure}[t]
    \centering
    \includegraphics[height=0.5\textheight,keepaspectratio]{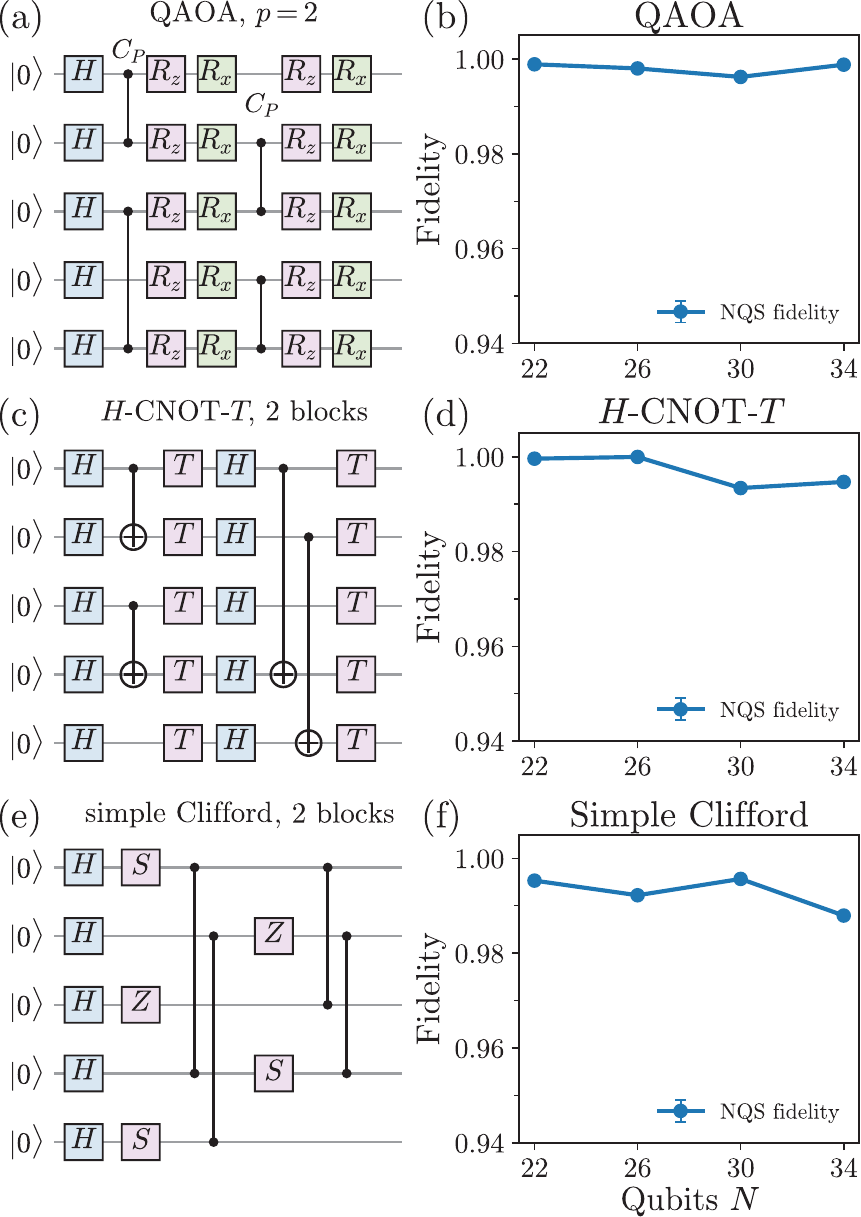}
    \caption{\textbf{Circuit-family tests.}
The same autoregressive Transformer statebank and training protocol are applied to \textbf{a,b}, QAOA with $p=2$; \textbf{c,d}, a two-block \HCNOTT{} circuit; and \textbf{e,f}, a two-block Clifford circuit. Left panels define the circuit families, and right panels report the final-state fidelity as a function of system size.}
    \label{fig:families}
\end{figure}

\section*{Universality}

The method does not rely on the special structure of the HCZCH benchmark. Fig.~\ref{fig:families} applies the same autoregressive Transformer architecture and the same statebank protocol to three additional circuit families. Here, a block denotes one repeated circuit unit. For the \HCNOTT{} family, one block is an all-qubit Hadamard layer, a CNOT layer, and an all-qubit $T$ layer. For the Clifford family, one block is a sparse $S/Z$ phase layer followed by a shifted CZ layer.

Fig.~\ref{fig:families}(a) and \ref{fig:families}(b) benchmark the quantum approximate optimization algorithm (QAOA) at $p=2$, a central variational circuit family for near-term optimization and quantum algorithm design~\cite{FarhiQAOA2014,FarhiQAOA2017,Hadfield2019QAOA,Zhou2020QAOA,Harrigan2021QAOA,Ebadi2022QAOA}. At generic angles, QAOA is non-Clifford, providing a natural test beyond stabilizer-based simulation. The statebank treats the diagonal cost layers as phase updates and the mixer layers as one-qubit local updates. The fidelity remains close to unity from $N=22$ to $34$.

Fig.~\ref{fig:families}(c) and \ref{fig:families}(d) test a two-block \HCNOTT{} circuit. This family is important because $H$, CNOT, and $T$ form a standard universal gate set~\cite{Barenco1995ElementaryGates,Boykin2000UniversalBasis,NielsenChuang2010,BravyiKitaev2005}. The same checkpointing procedure learns the branching induced by $H$ and the basis mixing induced by CNOT, while treating the $T$ layers as diagonal phase updates.

Fig.~\ref{fig:families}(e) and \ref{fig:families}(f) use a purely Clifford circuit as a control. This is not a regime in which a neural simulator should be expected to outperform specialized stabilizer methods, because Clifford circuits admit efficient classical simulation~\cite{Gottesman1998Heisenberg,AaronsonGottesman2004,AndersBriegel2006,Gidney2021Stim}. The neural checkpoint nevertheless reaches high fidelity. Together, these tests show that the same architecture handles diagonal phases, non-diagonal mixers, controlled gates, universal Clifford+$T$ dynamics, and pure Clifford dynamics without circuit-specific redesign.

\begin{figure}[t]
    \centering
    \includegraphics[width=\linewidth]{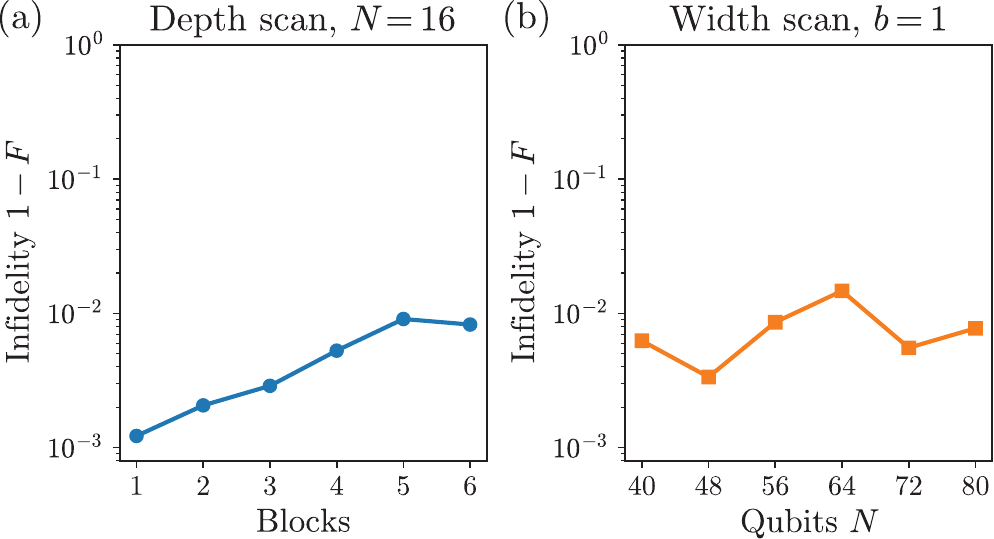}
    \caption{\textbf{Scaling.}
Infidelity in the Clifford scaling test. \textbf{a}, Depth scan at fixed width $N=16$ for $b=1,\ldots,6$ blocks, equivalently $D=2b$ learned layers. \textbf{b}, Width scan at fixed depth $b=1$ up to $N=80$. The stronger degradation with depth is consistent with accumulated layerwise approximation error.}
    \label{fig:scaling}
\end{figure}

\section*{Scaling}

As with any compressed simulator, the neural statebank requires more computational resources as the circuit becomes wider or deeper. These two directions stress the method in different ways. Width mainly increases memory use through model parameters, activations, batches, and optimizer state. Depth mainly affects accuracy, because each physical layer introduces another local learning step and approximation errors accumulate along the checkpoint trajectory.

Fig.~\ref{fig:scaling}(a) fixes $N=16$ and varies the number of blocks. The fidelity decays slowly with depth, consistent with accumulated approximation error rather than memory exhaustion. Fig.~\ref{fig:scaling}(b) fixes $b=1$ and varies $N$. At fixed depth, the infidelity remains at the $10^{-2}$ level even at $N=80$. The main cost of increasing $N$ is larger VRAM usage, not a systematic loss of accuracy. In these tests, the method is therefore limited more by depth than by width. Arbitrarily deep circuits would eventually require larger models, longer training, or additional error-control strategies. That is not the regime targeted here: uncorrected quantum hardware is itself restricted to moderate depths, where the statebank remains effective~\cite{Preskill2018,Bharti2022NISQ}.

\section*{Discussion}

The neural statebank provides a classical approach to quantum-circuit simulation in which the object being learned is not just the final state but the entire circuit trajectory. Rather than storing all $2^N$ amplitudes, it stores a sequence of Transformer checkpoints. Each checkpoint can sample from the learned Born distribution and evaluate $\psi(x)$ on requested bitstrings. The Transformer architecture supplies the capacity needed to represent highly entangled states~\cite{Vaswani2017,ZhangDiVentra2023,Viteritti2023}, while the autoregressive factorization enables direct sampling and substantially accelerates optimization~\cite{Sharir2020,HibatAllah2020,Luo2022ARTransformer,Barrett2022,IbarraGarciaPadilla2025AutoregressiveHubbard,Lange2024NQSReview}.

The statebank is not tied to a single favorable structure such as low entanglement, stabilizer simulability, or low magic. In the examples studied here, the same protocol applies to optimization circuits, circuits over the universal Clifford+$T$ gate set, and purely Clifford dynamics. On the HCZCH benchmark, it gives the best accuracy--resource trade-off among the tested approximate simulators that avoid exact state-vector storage in a regime combining long-range entanglement, magic, and non-diagonal branching. These results suggest that learning circuit trajectories as neural statebanks can complement existing classical simulation methods, especially for intermediate-scale circuits that are too structured to be fully generic but too mixed to fall within one specialized simulation regime.

\section*{Materials and Methods}

The following subsections specify the neural representation, layerwise losses, benchmark circuits, baseline simulators, and fidelity estimates used in the reported results.

\subsection*{Representation}

We use the notation of the main text. A computational-basis configuration is $x=(x_1,\ldots,x_N)$, and checkpoint $\theta_\ell$ represents the amplitude $\psi_{\theta_\ell}(x)$ after the first $\ell$ learned layers. Each checkpoint uses the autoregressive wavefunction in Eq.~\eqref{eq:nqs_ansatz}, so
\begin{equation}
    P_{\theta_\ell}(x)=|\psi_{\theta_\ell}(x)|^2
    =\prod_{i=1}^{N}p_{\theta_\ell}(x_i|x_{<i}) .
\end{equation}
The causal Transformer is fed the previous-bit token, $t_1=\START$ and $t_i=x_{i-1}$ for $i>1$, together with learned positional embeddings. Its output head gives two Born logits and two phase increments at each site. Amplitudes are evaluated by summing the selected log-probability and phase increments, and samples are generated directly by ancestral sampling from $P_{\theta_\ell}$.

For the circuits studied here, the initial all-Hadamard layer is represented exactly whenever it appears. This is achieved by setting the final output weights and biases to zero, giving $p_\theta(0|x_{<i})=p_\theta(1|x_{<i})=1/2$ and zero phase increment. The reported Transformer uses $d_{\rm model}=128$, two Transformer blocks, four attention heads, feed-forward dimension $d_{\rm ff}=256$, and zero dropout. The parameter counts are $268{,}932$ at $N=22$ and $270{,}468$ at $N=34$, with the increase mainly from positional embeddings.

Single-qubit rotations use $R_A(\alpha)=\exp(-i\alpha A/2)$ for $A\in\{X,Y,Z\}$. We use $T=\operatorname{diag}(1,e^{i\pi/4})$, $S=\operatorname{diag}(1,i)$, $Z=\operatorname{diag}(1,-1)$, and $\CP(\alpha)=\operatorname{diag}(1,1,1,e^{i\alpha})$. Controlled two-qubit gates are ordered as $|00\rangle,|01\rangle,|10\rangle,|11\rangle$, with the first qubit as the control:
\begin{equation}
    \CNOT|c,t\rangle=|c,t\oplus c\rangle,
    \qquad
    \CH=|0\rangle\langle0|\otimes I+|1\rangle\langle1|\otimes H .
\end{equation}
Global phases are not used in any loss or observable. Phase differences are always wrapped to $(-\pi,\pi]$ with $\operatorname{wrap}(\alpha)=\operatorname{atan2}(\sin\alpha,\cos\alpha)$.

\subsection*{Training objectives}

The exact trajectory obeys
\begin{equation}
    |\psi_\ell\rangle=U_\ell\cdots U_1|0\rangle^{\otimes N},
    \qquad
    \psi_\ell(x)=\langle x|\psi_\ell\rangle .
\end{equation}
The statebank approximates this trajectory by training one checkpoint at a time. At step $\ell$, $\psi_{\theta_{\ell-1}}$ is frozen and used as the source. The next checkpoint $\theta_\ell$ is trained to match the accessible teacher state
\begin{equation}
    \widetilde\psi_\ell=U_\ell\psi_{\theta_{\ell-1}},
\end{equation}
which is evaluated locally rather than constructed as a $2^N$-component vector.

For a diagonal layer, $U_\ell|x\rangle=e^{i\Phi_\ell(x)}|x\rangle$, so
\begin{equation}
    \widetilde\psi_\ell(x)=e^{i\Phi_\ell(x)}\psi_{\theta_{\ell-1}}(x).
\end{equation}
Samples are drawn from a mixture $\rho_\ell(x)=(1-\eta_\ell)|\psi_{\theta_{\ell-1}}(x)|^2+\eta_\ell2^{-N}$. In the HCZCH benchmark runs, the diagonal phase loss uses the unwrapped residual
\begin{equation}
    \Delta_\ell(x)=
    \Phi_{\theta_\ell}(x)-\Phi_{\theta_{\ell-1}}(x)-\Phi_\ell(x).
\end{equation}
We remove the unobservable global phase by subtracting the minibatch mean over the training distribution,
\begin{equation}
    \overline\Delta_\ell=\mathbb E_{\rho_\ell}\Delta_\ell(x),
    \qquad
    \delta_\ell(x)=\Delta_\ell(x)-\overline\Delta_\ell .
\end{equation}
The diagonal loss used in production is
\begin{equation}
    \mathcal L_{\rm diag}=\mathbb E_{\rho_\ell}\left[
    \bigl(\log|\psi_{\theta_\ell}(x)|-\log|\psi_{\theta_{\ell-1}}(x)|\bigr)^2
    +4\delta_\ell(x)^2
    \right].
\end{equation}
For a product of diagonal gates, $\Phi_\ell(x)$ is the sum of gate phases, for example $T_i: \pi x_i/4$, $S_i: \pi x_i/2$, $Z_i: \pi x_i$, $\CZ_{ij}: \pi x_ix_j$, and $\CP_{ij}(\alpha): \alpha x_ix_j$. For some Clifford diagnostics, we used a circular variant in which the mean phase is replaced by $\operatorname{Arg}\mathbb E_{\rho_\ell}e^{i\Delta_\ell(x)}$ and the phase loss by $\mathbb E_{\rho_\ell}[1-\cos\delta_\ell(x)]$.

For a non-diagonal local gate $u$ acting on active qubits $S$, write the spectator bits as $y=x_{\bar S}$ and the active bits as $\alpha=x_S$. For each sampled spectator configuration $y$, all active configurations are enumerated and the local teacher amplitudes are
\begin{equation}
    \widetilde\psi_\ell(y,\alpha)
    =\sum_\beta u_{\alpha\beta}\,
    \psi_{\theta_{\ell-1}}(y,\beta).
\end{equation}
The new checkpoint is compared with these teacher amplitudes on the same active-qubit block. We define
\begin{equation}
    M_y^\star=\sum_\alpha |\widetilde\psi_\ell(y,\alpha)|^2,
    \qquad
    M_y^\theta=\sum_\alpha |\psi_{\theta_\ell}(y,\alpha)|^2,
\end{equation}
and the conditional probabilities
\begin{equation}
    p_y^\star(\alpha)=\frac{|\widetilde\psi_\ell(y,\alpha)|^2}{M_y^\star},
    \qquad
    q_y^\theta(\alpha)=\frac{|\psi_{\theta_\ell}(y,\alpha)|^2}{M_y^\theta} .
\end{equation}
The KL divergence below is evaluated over the active configurations, $\operatorname{KL}(p\|q)=\sum_\alpha p(\alpha)\log[p(\alpha)/q(\alpha)]$. The phase and overlap terms are
\begin{equation}
    \mathcal L_\phi(y)=\sum_\alpha p_y^\star(\alpha)
    \left[1-\cos\bigl(\arg\psi_{\theta_\ell}(y,\alpha)-\arg\widetilde\psi_\ell(y,\alpha)\bigr)\right]
\end{equation}
and
\begin{equation}
    O_y^2=\left|\sum_\alpha
    \frac{\widetilde\psi_\ell(y,\alpha)^*\psi_{\theta_\ell}(y,\alpha)}{\sqrt{M_y^\star M_y^\theta}}
    \right|^2 .
\end{equation}
The local-gate loss used in the reported simulations is
\begin{align}
    \mathcal L_{\rm local}
    =\mathbb E_y\bigg[
    &4\,\operatorname{KL}(p_y^\star\|q_y^\theta)
    +4\mathcal L_\phi(y) \notag\\
    &+4\bigl(\log M_y^\theta-\log M_y^\star\bigr)^2
    +(1-O_y^2)
    \bigg].
\end{align}
We use $|S|=1$ for $H$, $R_X$, and $R_Y$ updates, and $|S|=2$ for CNOT and CH updates. CNOT and CH are trained with this local active-qubit objective, not as global bit-permutation regression.

\subsection*{Circuit definitions}

The HCZCH benchmark uses even $N$ and $h=N/2$. With qubits $q_0,\ldots,q_{N-1}$, define
\begin{equation}
    G_o=\left\{G\left(q_a,q_{h+((a+o)\bmod h)}\right):a=0,\ldots,h-1\right\} .
\end{equation}
For CH and CNOT, the lower-half qubit is the control. The production circuit is
\begin{equation}
\begin{aligned}
    H_{\rm all}&\rightarrow \CZ_4 \rightarrow T_{\rm all}\rightarrow \CH_5\\
    &\rightarrow T_{\rm all}\rightarrow \CNOT_7.
\end{aligned}
\end{equation}

The QAOA test has $p=2$ and starts from $H_{\rm all}$. The graph is $E=E_{\rm ring}\cup E_{\rm chord}$, with
\begin{align}
    E_{\rm ring}
    &=\{(i,i+1\bmod N): i=0,\ldots,N-1\},\\
    E_{\rm chord}
    &=\{(i,i+o\bmod N): i=0,\ldots,N-1\},
\end{align}
where $o=\max(2,\lfloor N/4\rfloor)$. For rounds $r=0,1$, the angles are
\begin{align}
    \gamma_r &= 0.08[0.90+0.12\sin(0.73(r+1))],\\
    \beta_r  &= 0.08[0.60+0.08\cos(0.41(r+2))].
\end{align}
Each round applies
\begin{equation}
    \prod_{(a,b)\in E}\CP_{ab}(-2\gamma_r)
    \rightarrow
    \prod_i R_{Z,i}(\gamma_r d_i)
    \rightarrow
    \prod_i R_{X,i}(2\beta_r),
\end{equation}
where $d_i$ is the graph degree.

The \HCNOTT{} test has two blocks. One block consists of an all-qubit Hadamard layer, a CNOT layer, and an all-qubit $T$ layer:
\begin{equation}
    H_{\rm all}\rightarrow C_0\rightarrow T_{\rm all}\rightarrow H_{\rm all}\rightarrow C_1\rightarrow T_{\rm all} .
\end{equation}
The first CNOT layer is $C_0=\{\CNOT(q_{2k},q_{2k+1}):k=0,\ldots,h-1\}$. The second uses $o=\lfloor N/4\rfloor$ and $C_1=\{\CNOT(q_k,q_{k+o}):k=0,\ldots,o-1\}$.

The simple Clifford test starts from $H_{\rm all}$ and consists of $b$ blocks. Each block contains a sparse $S/Z$ phase layer followed by a shifted lower-to-upper-half CZ layer:
\begin{equation}
H_{\rm all}
\rightarrow P_0\rightarrow B_0
\rightarrow P_1\rightarrow B_1
\rightarrow \cdots
\rightarrow P_{b-1}\rightarrow B_{b-1}.
\end{equation}
For block $r$, the sparse phase layer is
\begin{equation}
P_r=\prod_{i=0}^{N-1} P_{r,i},
\qquad
P_{r,i}=\begin{cases}
S_i, & (i+r)\bmod 4=0,\\
Z_i, & (i+r)\bmod 4=2,\\
I_i, & \mathrm{otherwise},
\end{cases}
\end{equation}
and the entangling layer is the shifted lower-to-upper-half CZ layer
\begin{equation}
B_r=
\left \{
\CZ\left(q_a,q_{h+((a+r)\bmod h)}\right):
 a=0,\ldots,h-1
\right \}.
\end{equation}

\subsection*{Benchmarks and fidelity}

Exact state-vector baselines use \qsim{}-style evolution with complex64 amplitudes. The raw state-array memory is $8\times2^N$ bytes, or $128\,\GiB$ at $N=34$. MPS baselines use \quimb{} tensor-network routines. Tensor-network contraction baselines use \quimb{} with \cotengra{} path optimization. The near-Clifford baseline is an optimized C++ implementation of Bravyi-style Clifford perturbation, including the $T$-block technique and low-level acceleration.

When an exact reference wavefunction is available, the reported fidelity is
\begin{equation}
    F(\psi_\ell,\psi_{\theta_\ell})=|\langle\psi_\ell|\psi_{\theta_\ell}\rangle|^2
    =\left|\sum_x \psi_\ell(x)^*\psi_{\theta_\ell}(x)\right|^2 .
\end{equation}
The plotted quantity is $1-F$. When direct summation is unavailable, we estimate the same overlap using reference samples $x^{(m)}\sim |\psi_\ell(x)|^2$:
\begin{equation}
    \widehat F=\left|\frac1n\sum_{m=1}^{n}
    \frac{\psi_{\theta_\ell}(x^{(m)})}{\psi_\ell(x^{(m)})}
    \right|^2 .
\end{equation}
This estimator is used only for evaluation. Its sampling error is shown in all figures, although it is typically smaller than the marker size. It is not used as a training loss or for checkpoint selection.

\section*{Acknowledgments}

We especially thank Michael Zaletel for discussions on existing classical simulation methods. We also thank Yi-Zhuang You for discussions. T.W. is grateful for support from the Harvard Quantum Initiative Fellowship. L.F. is supported by a Simons Investigator Award from the Simons
Foundation.

\bibliography{main}

\end{document}